# Absorbing Backside Anti-reflecting Layers for high contrast imaging in fluid cells.

Dominique Ausserré*[1], Refahi Abou Khachfe[2], Claude Amra[3] and Myriam Zerrad[3]

(1) "Molecular Landscapes, Biophotonic Horizons" group, UMR CNRS 6283, IMMM, Université du Maine, Avenue Olivier Messiaen, F-72000 Le Mans, France

(2) AUL University, South street, Jadra, Lebanon

(3) Institut Fresnel, Faculté des Sciences et Techniques de St Jérôme, 13397 Marseille Cedex 20

**Abstract:**

The single Anti-Reflecting (AR) layer is a classical problem in optics. When all materials are pure dielectrics, the solution is the so-called λ/4 layer. Here we examine the case of absorbing layers between non absorbing media. We find a solution for every layer absorption coefficient provided that the light goes from the higher towards the lower index medium, which characterizes backside layers. We describe these AR absorbing (ARA) layers through generalized index and thickness conditions. They are most often ultrathin, and have important applications for high contrast imaging in fluid cells.

The use of an anti-reflecting (AR) $\lambda/4$ layer in order to increase the optical contrast of an object in reflecting light microscopy is a usual and powerful trick. The first historical example of such a practice is probably the visualization of molecular steps reported by Langmuir and Blodgett in 1937 [1]. Recent examples are the study of alkane layers by Riegler and al. [2, 3] and the visualization of graphene layers by Novoselov and al. [4, 5]. They illustrate the importance of the method for current research issues. If $I$ is the reflected intensity of the object to visualize and $I_b$ that of the bare support, the contrast $C = (I - I_b)/(I + I_b)$ of the object is 100% when $I_b = 0$. This condition defines an AR surface. In the simplest case, it is achieved by coating a solid with a single layer. Analytical solutions for single dielectric AR layers illuminated at normal incidence are among the most famous results of the electromagnetic theory [6, 7]. They resume in two equations, which we will refer to as the classical results. The first one links the layer refractive index $n_1$ to the refractive indices $n_0$ and $n_3$ of the incident and emergent semi-infinite media through the relationship:

$$n_1^2 = n_0 n_3 \tag{1}$$

It warrants that the same amount of light is reflected by the two interfaces of the layer. The second equation is the thickness condition. It determines a thickness $e_1$ of the AR layer such that the addition of the beams multiply reflected at the two layer surfaces generates a destructive interference. The solution is periodic, the smallest value obeying the relationship $n_1 e_1 = \lambda/4$, from where the appellation of a $\lambda/4$ layer:

$$n_1 e_1 = (2p + 1)\lambda/4 \qquad (p \text{ integer}) \tag{2}$$

Equations (1) determines the refractive index of the AR layer in a unique way, which is unaffected by exchanging $n_0$ and $n_3$. The value of $n_1$ given by Equation (1) may correspond to exotic materials. At the glass-air interface for instance, it leads $n_1 \cong 1.27$, which requires the use of composite materials such as aero-gels. Here we demonstrate that when considering optically absorbing instead of dielectric materials, an infinite number of AR layers may be obtained on the sole condition that the refractive index of the incident medium is higher than that of the emergent medium. This configuration corresponds to backside layers. It is the one to consider when using an inverted microscope in reflecting light. This is illustrated in Figure 1b, while symbols are recalled in Figure 1a. It is especially useful in biosensor applications because the capturing events occur at the solid-liquid interface.

We assume the light monochromatic and its coherence length much larger than $\lambda$. The complex amplitude reflection coefficient of the solid (0) covered on the backside by layer (1) facing the output medium (3) is given by the Airy formula [6, 7]:

$$r_{013} = \frac{r_{01} + r_{13} e^{-2j\beta_1}}{1 + r_{01} r_{13} e^{-2j\beta_1}}, \tag{3}$$

where $r_{ij}$ holds for the Fresnel coefficient at the $ij$ interface. Its expression is $r_{ij}^{(p)} = (n_j \cos\theta_i - n_i \cos\theta_j)/(n_j \cos\theta_i + n_i \cos\theta_j)$ or $r_{ij}^{(s)} = (n_i \cos\theta_i - n_j \cos\theta_j)/(n_i \cos\theta_i + n_j \cos\theta_j)$, depending if the polarization (supposed linear) is $p$ (or TM) or $s$ (or TE). $\beta_1 = 2\pi n_1 e_1 \cos\theta_1/\lambda$ is the phase factor associated to layer (1). $\lambda$ is the light wavelength in vacuum, and $e_j$, $n_j$ and $\theta_j$ are the thickness, refractive index and refracted angle in medium $j$. We assume isotropic media and we adopt the sign conventions of Azzam & Bashara [7-9]. A single AR layer is obtained by setting $r_{013} = 0$ in equation (3), applied to the two elementary polarizations. It makes a system that we may call the AR equations. With purely dielectric media, $r_{01}$, $r_{13}$ and $\beta_1$ are real numbers, hence $e^{-2i\beta_1} = \pm 1$. With a positive sign, we obtain the so-called $\lambda/2$ layers, defined by $(2\pi/\lambda)n_1 e_1 \cos\theta_1 = m\pi$ ($m$ integer), or $e_1 = m\lambda/(2n_1 \cos\theta_1)$, which may exist only when the two semi-infinite media are identical. With a negative sign, we obtain the so-called $\lambda/4$ layers which we already mentioned. When the contrast layer 1 is absorbing, $n_1$, $\cos\theta_1$ and $\beta_1$ become complex quantities, and we note them in bold. We set $\boldsymbol{n_1} = n_1 - jk_1$, where $n_1$ and $k_1$ are real quantities. When the layer is sandwiched between two different materials, the AR equations impose : $\boldsymbol{\theta_1} = 0$ (Detailed derivations are given in supplementary materials SM1 [10]). Given two arbitrary semi-infinite media, there is no way to design a single anti-reflecting layer for a non normal incidence, even when using absorbing materials. At normal incidence, one cannot differentiate the two polarizations and the two AR equations merge into a single one:

$$\boldsymbol{n_1}^2 - j\frac{(n_3 - n_0)}{\tan\boldsymbol{\beta_1}}\boldsymbol{n_1} - n_0 n_3 = 0 \qquad (4)$$

Everyone is not familiar with the Fresnel formalism presently adopted. In particular, people in the optical multilayer community use to work with the Macleod formalism [11], which leads to a different form of equation (4), as explained in Supplementary Materials SM1 [10]. Equation (4) is transcendental. We cannot solve it in general. However, it remains possible in the frame of two approximations, corresponding to the limits of strongly absorbing (high k) and weakly absorbing (low k) materials, while it is assumed that the two surrounding semi-infinite media are pure dielectrics.

When considering strongly absorbing layers, we may suppose that $e_1 \ll \lambda$, hence $|\boldsymbol{\beta_1}| \ll 1$, because no light could go through thick layers. We will check this afterwards. Up to second order in $\boldsymbol{\beta_1}$, equation (4) takes the simple form $\boldsymbol{n_1}^2/n_0 n_3 - j(1 - n_0/n_3)\lambda/2\pi e_1 - 1 = 0$. Separating real and imaginary parts, we get:

$$\begin{cases} n_1^2 \cong n_0 n_3 + k_1^2 \\ e_1 \cong \frac{(n_0/n_3 - n_3/n_0)}{n_1 k_1} \frac{\lambda}{4\pi} \end{cases} \qquad (5\text{-}6)$$

Since $e_1$ in Equation (6) is real and positive, the existence of the low thickness solution requires $n_0 > n_3$. Therefore, the highly absorbing AR layer exists only for the inverted geometry evoked in Fig. 1b. Typical values for $n_0$ and $n_3$ in a biosensor experiment are 1.52 (for glass) and 1.34 (for water solution), showing that $e_1$ is a few nanometers. This confirms our initial assumption. It is remarkable, and a priori fortuitous since it was derived under a low $e_1$, hence high $k_1$ assumption, that equation (5) reduces to Equation (1) when $k_1$ vanishes. In Figure 2, we used the reduced parameters $\nu_1 = n_1/\sqrt{n_0 n_3}$ and $\kappa_1 = k_1/\sqrt{n_0 n_3}$ because Equation (5) is homogeneous with respect to $n_1$ and $k_1$. Figure 2a compares the optimal values $\nu_1(\kappa_1)$ given by equation (5) and the exact result obtained by a numerical approach (how we proceed is explained in supplementary material SM2 [12]). In Figure 2b, Equation (5) is shown to be valid within a 0.25 % error at any $\kappa_1$ value, that is to say for arbitrarily absorbing materials. The unique previously known solution $\nu_1 = 1$ is marked in the figure by a red circle. The line $\kappa_1 = \nu_1$ in Fig. 2a describes the asymptotic behavior when both quantities become very high. Notice that $k_1 < n_1$ is a necessary condition for the existence the AR absorbing (ARA) layer, which should reduce the role of metals compared to semi-conductors [6]. However, optical properties of ultrathin metal films differ from those of the bulk material. This is the case for instance with gold ultrathin films, which optical properties become compatible with equations 5-6 at very low thickness [13], or when made of assembled nanoparticles [14]. Other convenient materials are non stoechiometric metal oxides, polymer or glass films with embedded colorants, pigments, metal or oxide nanoparticles, or any material with a natural absorption band at the working wavelength. Last but not least, equations (5-6) are satisfied at 450 nm with a stack of 11 or 12 graphene monolayers on a glass substrate in water. This case will be developed in a further work.

When $k_1 = 0$, $e_1$ should also reduce to the classical result $\lambda/4n_1$, but this is not predicted by Equation (5). By contrast with the index condition, the equation obtained in the high $k_1$ regime is not valid for any layer absorption coefficient. We must now focus on the other asymptotic regime, where $k_1$ is arbitrary small. Then, we can derive $e_1$ by using a different approximation, namely by setting $\boldsymbol{\beta_1} = \pi/2 - \boldsymbol{\varepsilon_1}$, where $\boldsymbol{\varepsilon_1} = \pi/2(1 - 4\boldsymbol{n_1}e_1/\lambda)$. The module of $\boldsymbol{\varepsilon_1}$ is small. Equation (2) leads, up to second order in $k_1$:

$$\begin{cases} n_1^2 \simeq n_0 n_3 + \frac{\pi}{2}(n_0 - n_3)k_1 - 3k_1^2 + o(k_1^3) \\ e_1 \simeq \frac{\lambda}{4n_1}\left\{1 - \frac{4}{\pi(n_0-n_3)}k_1 + \frac{k_1^2}{n_0 n_3} + o(\kappa_1^3)\right\} \end{cases} \quad (7\text{-}8)$$

Derivation of Equations (7-8) is reported in Supplementary Materials SM3 [15]. Low k anti-reflecting layers were previously envisaged by Moiseev and Vinogradov [16], but imposing the air as the incident medium. Equation (7) conforms to their result. Writing Equation (8) in terms of $k_1$ with the help of Equation (7), we get, at first order in $k_1$:

$$e_1 = \left(\lambda/4\sqrt{n_0 n_3}\right)\{1 - k_1[(\pi/4)(n_0 - n_3)/n_0 n_3 + (4/\pi)/(n_0 - n_3)]\} \quad (9)$$

When $n_0 < n_3$ the optimal thickness $e_1$ is an increasing function of the absorption coefficient $k_1$. The anti-reflecting layer becoming at once thicker and more absorbing with increasing $k_1$, a point is rapidly attained where the light cannot go through, and the destructive interference between the beams reflected by the two interfaces of the layer cannot take place. Figures 2c and 2d compare, with different ranges of $\kappa_1$ values (0 to 2 and 0 to 0.1, respectively), the optimal trajectories $\nu_1(\kappa_1)$ obtained according to equation (5), to equation (7) up to first order, to equation (7) up to second order, and to numerical calculation. Figure 2d focuses on the low $\kappa_1$ region. The second order low $\kappa_1$ expansion breaks down as soon as $\kappa_1$ becomes larger than a few per cent, because of the presence of an inflexion point on the numerical curve.

The ARA thickness given by equation (6) and by equation (8) is very different. For the sake of generality, we introduce a third reduced variable $\delta_1 = (2\pi n_0/\lambda)e_1$, that we may call the dimensionless layer thickness. Figure 3a and 3b compare with different $\kappa_1$ ranges the values of $\delta_1$ obtained from these equations and from numerical calculation. The approximate forms of the thickness condition in the two extreme regimes $\kappa_1 \gg 1$ and $\kappa_1 \ll 1$ strongly differ from the numerical result in the intermediate regime. However, we are able to merge the two asymptotic solutions in a single semi-empirical expression which covers the full range of isotropic AR layers from purely dielectric to highly absorbing. Under the control of numerical calculation, we obtained:

$$e_1 \cong \frac{\lambda}{4\pi} \frac{(n_0-n_3)}{n_1 k_1}\left[1 - e^{-\frac{\pi k_1}{(n_0-n_3)}}\right] \quad (10)$$

This expression reduces to equation (6) when $k_1$ is large and identifies with equation (8) when $k_1$ is small. The graph of the function defined by equation (10) is also displayed in Fig. 3. The difference with the numerical solution never exceeds 3.5%. Figures 4a and 4b display the reflectance $R = |r_{013}|^2$ of backside layers with various $\kappa_1$ values, all satisfying the index condition, as a function of their reduced thickness $\delta_1$ (4a) or inverted reduced thickness $1/\delta_1$ (4b), which we may call as well the reduced wavelength. The value of $n_0/n_3$ is fixed to 1.14. It corresponds to a water output medium, when using a glass window as a transparent solid support as in the setup displayed in Fig. 1b. In a given frame, each curve corresponds to a different value of $\kappa_1$, comprised between 0 and 2, the optimal value $\nu_1(\kappa_1)$ being consequently determined by equation (5). The yellow curves correspond to the dielectric case. Each curve presents a zero reflectance when the thickness condition , equation (7), is satisfied. The existence of multiple $\{\kappa_1, \nu_1, \delta_1\}$ solutions for a fixed value of $n_0/n_3$ contrasts with the uniqueness of the dielectric AR layer solution. Absorbing materials bring a high flexibility in the design of AR layers. For instance, making an AR layer on glass in air, which we said is difficult with dielectric materials, becomes simple (on the backside of the glass) with absorbing materials. One may simply impose $n_1 = 1.52$ and it results in $k_1 = 0.89$. Such a layer may be achieved by exposing the top layer of the glass surface to ionic beams [17]. Fig. 4b shows that ARA layers act as edge filters on the reflected signal. This may be used for generating color changes in biosensor applications. Figures 4c and 4d show the absorbance of the same layers [6, 7] as a function of, respectively, $\delta_1$ and $1/\delta_1$. There is no marked absorption peak when the reflectivity vanishes, showing that extinction is essentially an interferential effect [15]. To resume, ARA layers are all described by equation (5) and equation (10). We may write those in terms of only four reduced parameters $\nu_1$, $\kappa_1$, $\delta_1$ and $n_0/n_3$. Since optimal values of $\nu_1$ and $\delta_1$ are themselves functions of $\kappa_1$ and $n_0/n_3$, the two latter parameters are sufficient to scan the full AR layer space. Considering the use of ARA layers for contrast enhancement in biosensing experiments, Figure 5a and 5b show how the contrast of an additional "test" layer 2 with thickness 1 nm and refractive index 1.46, inserted between the ARA layer deposited on a glass support and the water emergent medium ($n_0/n_3 = 1.14$), depends on the ARA layer thickness (Fig. 5a) or wavelength (Fig. 5b) around their optimum value, for the same set of $\kappa_1$ values as in Fig. 4, and using the same color code. It is remarkable that the more important is $\kappa_1$, the more asymmetric is the contrast curve. This is because the additional layer to visualize, supposed dielectric, cannot compensate a change in the AR thickness when the latter is absorbing. The ARA layers introduced herein will in principle allow to visualize arbitrarily thin objects at a solid/water and a solid/air interface with a 100% contrast. In wide filed imaging applications however, one should seek for a compromise between resolution and contrast because the AR conditions are valid for normal incidence. ARA layers can also be used without any compromise for high resolution scanning imaging with the help of a focused laser beam. Beside biosensing, they are expected to be useful whenever high contrast imaging on conducting surfaces is required. This concerns electrochemical imaging and a number of processes and devices in the semi-conductor industry.

**Acknowledgements**

This work was granted by ANR under PNANO-07-050 program.

**Figure caption**

**Figure 1 Geometry of the optical problem and symbols. a,** Index 1 holds for the AR layer. **b,** Example of a corresponding experimental set-up. The liquid flows above the AR layer. Capturing events at the liquid/layer interface restore a non zero reflectivity.

**Figure 2 The index condition for absorbing layers. a,** relationship between the real and imaginary parts of the absorbing AR layer in the $\{\kappa_1, \nu_1\}$ plane. The curves obtained according to (grey full line) equation (3) and from (red cruces) numerical calculation cannot be separated by the eye. The asymptotic behavior is marked by the bisector black line. All solutions are above that line, which shows that dielectric layers require that $\nu_1 > \kappa_1$ . The unique solution corresponding to the dielectric layer is marked by a red circle. All other solutions are new ones. **b,** per cent difference between the values

obtained numerically and those predicted by equation (3). **c,** optimal trajectories $\nu_1(\kappa_1)$ in the $\kappa_1$ range $\{0, 4\}$ obtained according to (grey line) equation (3), (light blue line) equation (5) up to first order, (green line) equation (5) up to second order, and (red cruces) numerical calculation. **d,** same in the $\kappa_1$ range $\{0, 0.1\}$.

**Figure 3 The thickness condition for absorbing layers. a,** Optimal reduced thickness $\delta_1$ as a function of the reduced absorption coefficient $\kappa_1$ according to (dark blue line) equation (4), (green line) equation (6), (light blue line) equation (7), and (red cruces) numerical calculation. **b,** Focus on the transition between the two asymptotic regimes, using the same color code.

**Figure 4 Reflectance and absorbance of a backside absorbing layer. a,** Reflectance R of a layer obeying the index condition equation (3). as a function of the reduced layer thickness $\delta_1$ for $n_0/n_3 = 1.14$ and for various $\kappa_1$ values ranging from 0 to 2: $\kappa_1 =$ (yellow) 0, (violet) 0.1, (light blue) 0.3, (green) 0.6, (red) 1, (dark blue) 2. **b,** same as a function of the reduced wavelength of light $1/\delta_1$. **c,** Absorbance of the backside layer as a function of $\delta_1$. **d,** Same as a function of $1/\delta_1$. By making the correspondence between each minimum in Fig. 4a with Fig. 4c, one can check that the absorbance of the optimal layer is always close to 0.1.

**Figure 5 Contrast of an object in the (sub)nanometer range. a,b,** The object has a thickness 1 nm and a refractive index 1.46 and is observed in water ($n_0/n_3 = 1.14$). The contrast is calculated as a function of (log scale): **a,** the reduced thickness $\delta_1$; **b,** the reduced wavelength $1/\delta_1$, for six different absorbing coefficients $\kappa_1$ ranging from 0 to 2.

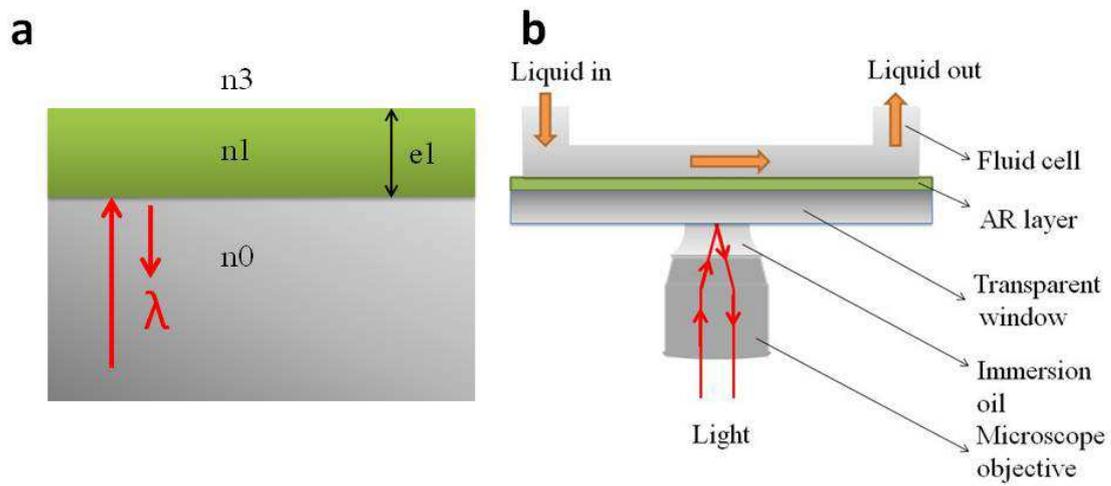

Figure 1

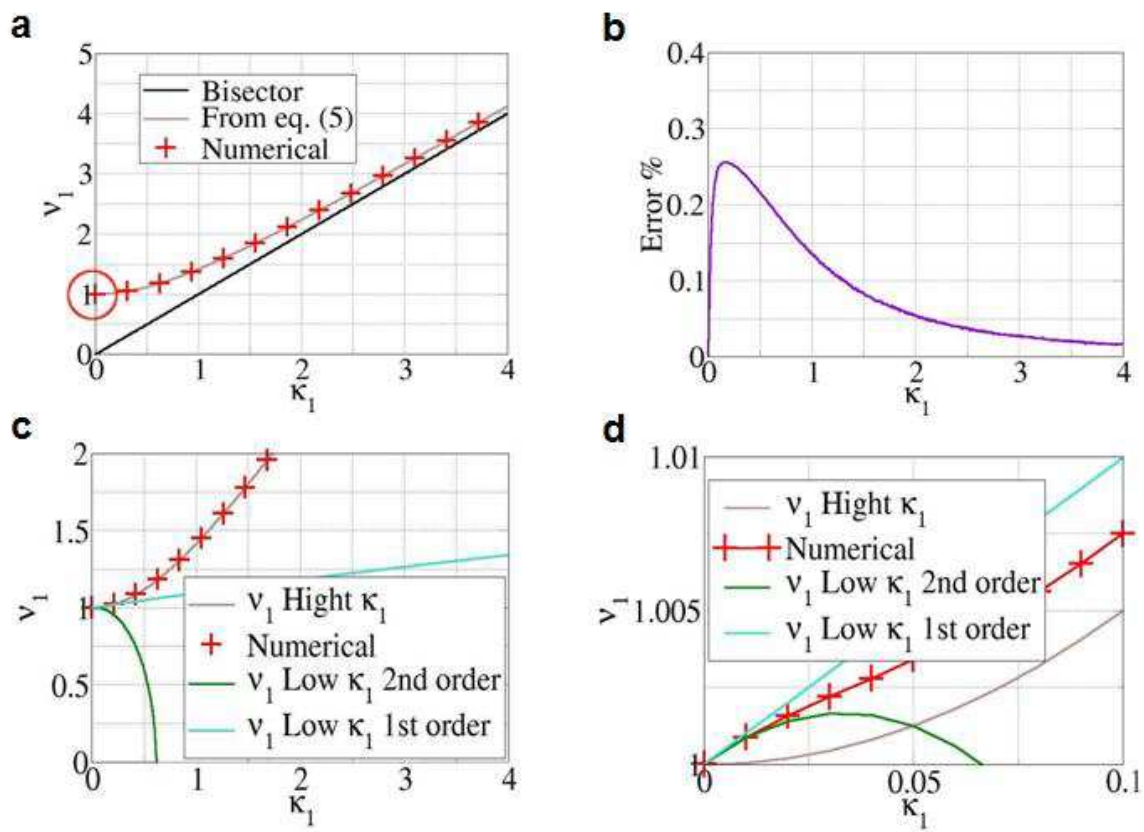

Figure 2

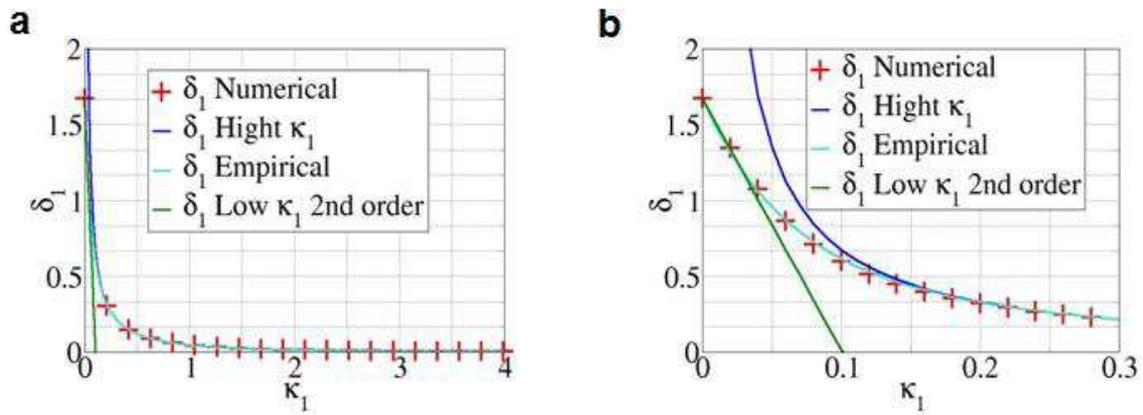

Figure 3

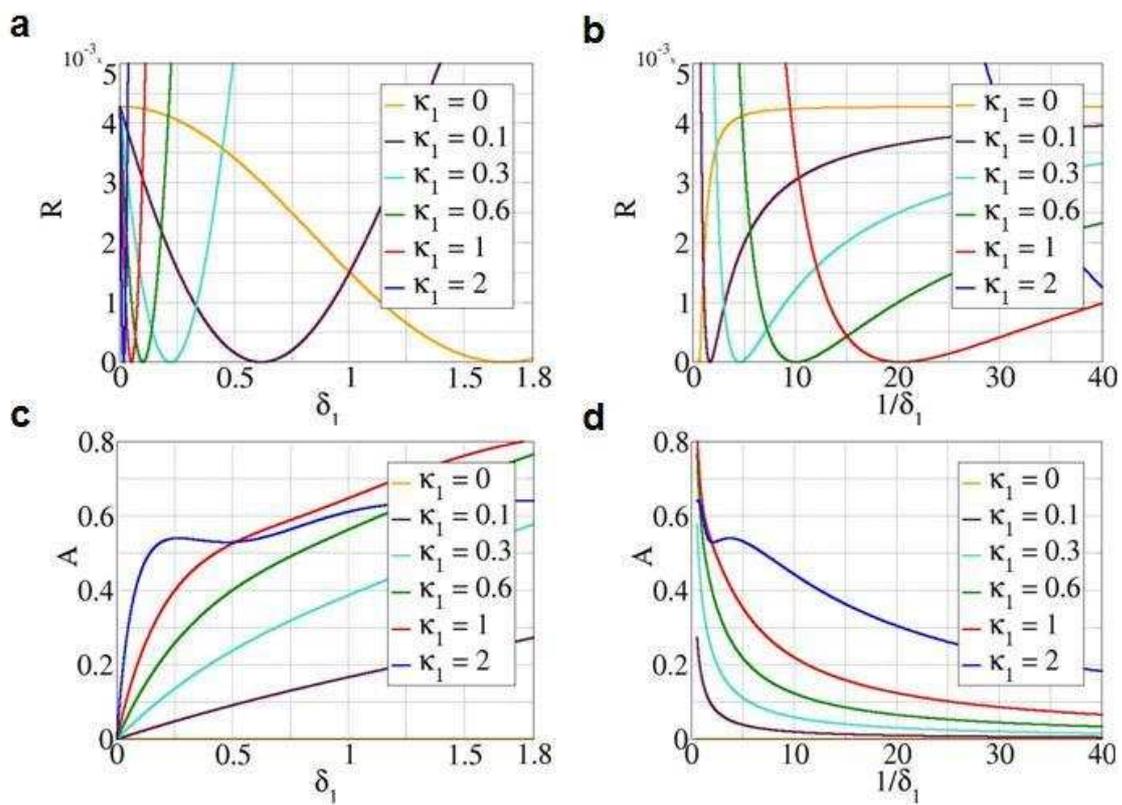

Figure 4

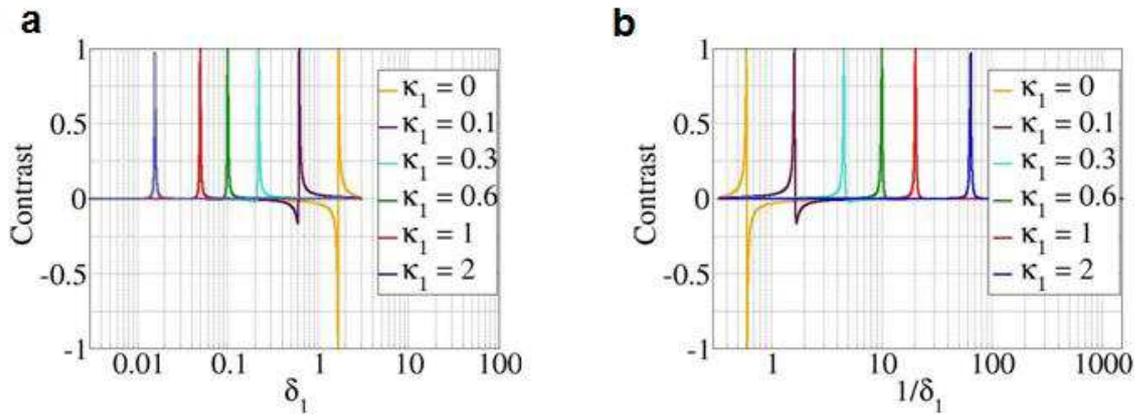

Figure 5

**Supplemental Material SM1**

**The AR layer problem**

A single AR layer is obtained by setting $r_{013} = 0$ in equation (3) applied to the two elementary polarizations. Assuming isotropic media and adopting the sign conventions of Azzam & Bashara [7-9], the Fresnel coefficients are respectively $r_{ij}^{(p)} = (n_j \cos\theta_i - n_i \cos\theta_j)/(n_j \cos\theta_i + n_i \cos\theta_j)$ and $r_{ij}^{(s)} = (n_i \cos\theta_i - n_j \cos\theta_j)/(n_i \cos\theta_i + n_j \cos\theta_j)$, depending if the polarization (supposed linear) is $p$ (or TM) or $s$ (or TE).

**A dielectric layer between dielectric media**

When all media are purely dielectric, $r_{01}$, $r_{13}$ and $\beta_1$ are real numbers, hence $e^{-2i\beta} = \pm 1$, and there are two families of solutions, given by the following systems :

$$\begin{cases} e^{-2j\beta_1} = 1 \\ r_{01} = -r_{13} \end{cases} \quad \text{(S1-S2)}$$

$$\begin{cases} e^{-2j\beta_1} = -1 \\ r_{01} = r_{13} \end{cases} \quad \text{(S3-S4)}$$

Eq.S1 leads to the so-called $\lambda/2$ layers, defined by $(2\pi/\lambda)n_1 e_1 \cos\theta_1 = m\pi$ ($m$ integer), or $e_1 = m\lambda/(2n_1 \cos\theta_1)$.

Introducing the notation $c_i = \cos\theta_i$, Eq. S2 reduces to $c_0 n_3 = c_3 n_0$ for $p$ polarization, and to $n_0 c_0 = n_3 c_3$ for $s$ polarization. Combining the two gives $n_0 = n_3$ and $c_0 = c_3$. Solutions may exist only when the two semi-infinite media are identical.

Eq. S3 corresponds the so-called $\lambda/4$ layers, defined by $(2\pi/\lambda)n_1 e_1 \cos\theta_1 = m\pi + \pi/2$ ($m$ integer), or $e_1 = \left(m + \frac{1}{2}\right)\lambda/(2n_1 \cos\theta_1)$.

Introducing Fresnel coefficients, Eq. S4 reduces to $n_1 c_0 n_1 c_3 = n_3 c_1 n_0 c_1$ for $p$ polarization, and to $n_0 c_0 n_3 c_3 = n_1 c_1 n_1 c_1$ for $s$ polarization. Combining $p$ and $s$ gives: $n_1^2 = n_0 n_3$

and $c_0 c_3 = c_1^2$. Setting $X = n_1 sin\theta_1$ (Snell invariant), the latter gives $(1 - X^2/n_0^2)(1 - X^2/n_3^2) = (1 - X^2/n_1^2)^2$ from where $n_2 = n_0 = n_1$ (no interface) or $X = 0$ (normal incidence). **Therefore, between two arbitrary dielectric media, the dielectric AR layer only exists for normal incidence**.

Therefore, the solution of Eq. S3 reduces to $n_1 e_1 = (2p + 1) \lambda/4$, $p$ being an integer.

**An absorbing layer between two dielectric media**

When layer 1 is absorbing, $r_{013} = 0$ imposes:

$$r_{01,s} r_{13,p} = r_{01p} r_{13,s} \tag{S5}$$

Expressing Fresnel coefficients, and using $X = \boldsymbol{n_1} sin\boldsymbol{\theta_1}$ (Snell invariant), Eq. S5 becomes:

$$\frac{X^2}{\boldsymbol{n_1}^2}(n_0^2 - \boldsymbol{n_1}^2)(n_2^2 - \boldsymbol{n_1}^2)\left(\frac{c_0}{n_2} - \frac{c_2}{n_0}\right) = 0$$

In order to have Eq. S5 true, at least one of the following conditions must be satisfied: $X = 0$ (normal incidence); $\boldsymbol{n_1}^2 = n_0^2$ (no layer); $\boldsymbol{n_1}^2 = n_2^2$ (no layer) or $n_0^2 = n_2^2$ (same semi-infinite medium on both sides). **We conclude that it is not possible to realize a single anti-reflecting layer for non normal incidences, even with absorbing materials**.

<u>**Using the Macleod formalism,**</u>

one would write [10]: $r_{013} = (n_0 - Y)/(n_0 + Y)$, where $Y$ is the complex admittance of the single layer. Using the same symbols as in the main text, it expresses as $Y = [n_3 - j\boldsymbol{n_1} tan\boldsymbol{\beta_1}]/[1 - j(n_3/\boldsymbol{n_1})tan\boldsymbol{\beta_1}]$. Setting $r_{013} = 0$ in equation (3), we get $\boldsymbol{n_1}^2 + j\frac{(n_3 - n_0)}{tan\boldsymbol{\beta_1}}\boldsymbol{n_1} - n_0 n_3 = 0$ instead of equation (4). The sign of the second term is inverted. This difference is simply due to a change in sign conventions. Equation (4) corresponds to writing $\boldsymbol{n_1} = n_1 - jk_1$, while the latter equation corresponds to writing $\boldsymbol{n_1} = n_1 + jk_1$.

Page 17

## Supplemental Material SM2 : Derivation of Equations 7 and 8

Introducing the reduced variable $\nu_1$, we write Equation (4) as:

$$\nu_1^2 + j\frac{(n_0/n_3 - 1)}{tan\beta_1}\sqrt{\frac{n_3}{n_0}}\nu_1 - 1 = 0$$

$\beta_1$ may be also be written in terms of the reduced parameters as:

$$\beta_1 = \frac{2\pi n_0 e_1}{\lambda}\frac{n_1}{n_0} = \frac{2\pi n_0 e_1}{\lambda}\frac{n_1}{\sqrt{n_0 n_3}}\sqrt{\frac{n_3}{n_0}} = \delta_1\sqrt{\frac{n_3}{n_0}}\nu_1$$

Setting:

$$\beta_1 = \pi/2 - \varepsilon_1$$

We have:

$$tan\beta_1 = 1/tan\varepsilon_1$$

Equation (4) becomes:

$$\nu_1^2 + j(n_0/n_3 - 1)\sqrt{\frac{n_3}{n_0}}\nu_1 tan\varepsilon_1 - 1 = 0$$

$$tan\varepsilon_1 \cong \varepsilon_1 = \pi/2 - \sqrt{\frac{n_3}{n_0}}\nu_1\delta_1$$

$\varepsilon_1 = 0$ corresponds to $\sqrt{\frac{n_3}{n_0}}\nu_1\delta_1 = \frac{\pi}{2}$, or $\delta_1 = \frac{\pi}{2\nu_{1D}}\sqrt{\frac{n_0}{n_3}}$, where we have set: $\nu_{1D} = \sqrt{n_0 n_3}$

Equation (4) becomes:

$$\nu_1^2 + j(n_0/n_3 - 1)\sqrt{\frac{n_3}{n_0}}\nu_1\left(\pi/2 - \sqrt{\frac{n_3}{n_0}}\nu_1\delta_1\right) - 1 = 0$$

Or else:

$$\nu_1^2\left[1 - j\left(\frac{n_3}{n_0}\right)(n_0/n_3 - 1)\delta_1\right] + j\frac{\pi}{2}(n_0/n_3 - 1)\sqrt{\frac{n_3}{n_0}}\nu_1 - 1 = 0$$

Separating real and imaginary parts, we get respectively:

$$\nu_1{}^2 - \kappa_1{}^2 - 1 - 2\left(\frac{n_3}{n_0}\right)\left(\frac{n_0}{n_3} - 1\right)\delta_1\nu_1\kappa_1 + \frac{\pi}{2}\left(\frac{n_0}{n_3} - 1\right)\sqrt{\frac{n_3}{n_0}}\kappa_1 = 0 \tag{SM3-1}$$

and:

$$-2j\nu_1\kappa_1 - j(\nu_1{}^2 - \kappa_1{}^2)\left(\frac{n_3}{n_0}\right)(n_0/n_3 - 1)\delta_1 + j\frac{\pi}{2}(n_0/n_3 - 1)\sqrt{\frac{n_3}{n_0}}\nu_1 = 0$$

$$\tag{SM3-2}$$

Rearranging SM3-2, we get:

$$\delta_1 = \frac{\pi}{2}\sqrt{\frac{n_0}{n_3}}\frac{\nu_1}{(\nu_1{}^2 - \kappa_1{}^2)}\left[1 - \frac{4}{\pi}\frac{\sqrt{\frac{n_0}{n_3}}}{(n_0/n_3 - 1)}\kappa_1\right]$$

Using:

$$\frac{\nu_1}{(\nu_1{}^2 - \kappa_1{}^2)} \cong \frac{1}{\nu_1}\left[1 + \left(\frac{\kappa_1}{\nu_1}\right)^2\right]$$

we obtain the expansion of $\delta_1$ up to second order in $\kappa_1$:

$$\delta_1 = \frac{\pi}{2}\sqrt{\frac{n_0}{n_3}}\frac{1}{\nu_1}\left[1 + \left(\frac{\kappa_1}{\nu_1}\right)^2\right]\left[1 - \frac{4}{\pi}\frac{\sqrt{\frac{n_0}{n_3}}}{(n_0/n_3 - 1)}\kappa_1\right]$$

, which we write:

$$\delta_1 = \frac{\pi}{2}\sqrt{\frac{n_0}{n_3}}\frac{1}{\nu_1} - \frac{2}{\nu_1}\frac{n_0/n_3}{(n_0/n_3 - 1)}\kappa_1 + \frac{\pi}{2}\sqrt{\frac{n_0}{n_3}}\frac{1}{\nu_1}\left(\frac{\kappa_1}{\nu_1}\right)^2$$

This is Equation (8) in the article.

_________________________________________________________________

We now inject this result in Eq. SM3-1 and get successively:

$$\frac{\nu_1{}^2 - \kappa_1{}^2 - 1}{\left(\frac{n_0}{n_3} - 1\right)\left(\frac{n_3}{n_0}\right)} + \frac{\pi}{2}\sqrt{\frac{n_0}{n_3}}\kappa_1 = \pi\kappa_1\sqrt{\frac{n_0}{n_3}}\left[1 - \frac{4}{\pi}\frac{\sqrt{n_0/n_3}}{(n_0/n_3 - 1)}\kappa_1 + \left(\frac{\kappa_1}{\nu_1}\right)^2\right]$$

$$\frac{\nu_1{}^2 - \kappa_1{}^2 - 1}{\left(\frac{n_0}{n_3} - 1\right)\sqrt{\frac{n_3}{n_0}}} = \frac{\pi}{2}\kappa_1\left[1 - \frac{8}{\pi}\frac{\sqrt{\frac{n_0}{n_3}}}{(n_0/n_3 - 1)}\kappa_1 + 2\left(\frac{\kappa_1}{\nu_1}\right)^2\right]$$

$$\nu_1{}^2 - \kappa_1{}^2 = 1 + \frac{\pi}{2}\left(\frac{n_0}{n_3} - 1\right)\sqrt{\frac{n_3}{n_0}}\kappa_1\left[1 - \frac{8}{\pi}\frac{\sqrt{\frac{n_0}{n_3}}}{(n_0/n_3 - 1)}\kappa_1 + 2\left(\frac{\kappa_1}{\nu_1}\right)^2\right]$$

The latter form is interesting because it highlights a correction to the high $\kappa_1$ equation (5) in the article.

Isolating $v_1^2$, we obtain :

$$v_1^2 = 1 + \frac{\pi}{2}\left(\frac{n_0}{n_3} - 1\right)\sqrt{\frac{n_3}{n_0}}\kappa_1 \left[1 - \frac{8}{\pi}\frac{\sqrt{\frac{n_0}{n_3}}}{(n_0/n_3 - 1)}\kappa_1 + \frac{1}{\frac{\pi}{2}\left(\frac{n_0}{n_3}-1\right)\sqrt{\frac{n_3}{n_0}}}\kappa_1 + 2\left(\frac{\kappa_1}{v_1}\right)^2\right]$$

, and finally:

$$v_1^2 = 1 + \frac{\pi}{2}\left(\frac{n_0}{n_3} - 1\right)\sqrt{\frac{n_3}{n_0}}\kappa_1 - 3\kappa_1^2$$

This is equation (7) in the article.

# Supplementary Material SM3

**Reflectivity coefficient**

The total reflectivity is $R = |r_{013}|^2$ and $r_{013}$ is given by Equation (3).

For a normal incidence we have $r_{01} = \frac{n_0 - n_3}{n_0 + n_1}$, $r_{13} = \frac{n_1 - n_3}{n_1 + n_3}$ and $\beta_1 = \frac{4\pi n_1 e_1}{\lambda}$

**Transmission Coefficient**

The transmission coefficient is $T = \frac{n_3}{n_0} * |t_{013}|^2$ with $t_{013} = \frac{t_{01} * t_{13} e^{-2j\beta_1}}{1 + r_{01} r_{13} e^{-2j\beta_1}}$

For a normal incidence we have $t_{01} = \frac{2n_0}{n_0 + n_1}$ and $t_{13} = \frac{2n_1}{n_1 + n_3}$

**Absorption Coefficient**

The absorption coefficient is $A = 1 - T - R$

**Numerical solution of equation (4) of the article**

Equation (4) corresponds to $|r_{013}| = 0$, and therefore to the minimum of $|r_{013}(\nu_1, \kappa_1, \delta_1)|$.

For every fixed value of $\kappa_1$, we use the simplex optimization method in order to determine the optimal values $\nu_{1opt}(\kappa_1)$ and $\delta_{1opt}(\kappa_1)$ such as $|r_{013}(\nu_1, \kappa_1, \delta_1)|$ is minimum, and we check that this minimum value is zero. The minimization is performed under Matlab environment.